\documentstyle[amsfonts,12pt]{article}
\def\one{1\hskip-.37em 1}

\def\half{\textstyle{\frac{1}{2}}}

\def\H{{\cal H}}
\def\D{{\cal D}}
\def\De{\Delta}

\def\ra{\rightarrow}
\def\tint{{\textstyle\int}}

\def\s{\hskip.08em}
\def\d{\partial}
\def\o{\overline}

\def\b{\begin{eqnarray*}}     
\def\e{\end{eqnarray*}}       
\def\bn{\begin{eqnarray}}     
\def\en{\end{eqnarray}}       
\def\<{\langle}
\def\>{\rangle}

\def\{{\lbrace}
\def\}{\rbrace}
\begin{document}
\title{Phase Space Geometry in \\Classical and Quantum Mechanics
\footnote{Presented at the Second International Workshop on Contemporary 
Problems in Mathematical Physics, Cotonou, Benin, October 28-November 2,
2001}}
\author{John R. Klauder
\footnote{Electronic mail: klauder@phys.ufl.edu}\\
Departments of Physics and Mathematics\\
University of Florida\\
Gainesville, FL  32611}
\date{}     
\maketitle
\begin{abstract}
Phase space is the state space of classical mechanics, and this manifold 
is normally endowed only with a symplectic form. The geometry of quantum 
mechanics is necessarily more complicated. Arguments will be given to 
show that augmenting the symplectic manifold of classical phase space 
with a Riemannian metric is sufficient for describing quantum mechanics. 
In particular, using such spaces, a fully satisfactory geometric version 
of quantization will be developed and described.
\end{abstract}
\section{Introduction}
What is the difference between classical mechanics and quantum mechanics? 

{}From a certain perspective, it is surely true that the differences appear 
to be vast. One theory is deterministic, the other is stochastic. One 
theory involves point particles and their trajectories, the other 
involves wave functions generally spread out over space. One theory 
involves commuting algebraic expressions, the other involves generally 
noncommuting algebraic expressions corresponding, in each case, to 
observables. These differences are indeed vast and well known, and 
of course they are all true. In brief, if one wants to find big 
differences, then it is not too difficult to do so. However, let us 
take a different perspective.

Instead of focusing on vast distinctions, let us see how close we can 
bring the formulations of classical and quantum mechanics to each other. 
If we can bring them close to each other, then we may be able to shed 
comparative light on one side from the vantage point of the other side, 
a procedure which may be rather useful in gaining further understanding 
of both theories.

Prior to the discovery of quantum mechanics, there was a wealth of 
significant developments in the arena of classical mechanics. The range of 
applicability of classical mechanics is enormous, and, for many sorts of 
problems, a classical description is frequently all that is needed. With 
the advent of quantum mechanics, a wider family of problems may be 
successfully addressed. Sometimes one reads that is is necessary to let 
$\hbar$, Planck's constant$/2\pi$, go to zero, i.e., $\hbar\ra0$, if one 
wishes to describe classical mechanics. Our viewpoint is quite the 
opposite in that $\hbar$ is, after all, not zero in the real world, and 
it should be accepted for the nonzero value that it has, 
$\hbar\simeq 10^{-27}$ erg-sec. Consequently, it should not be a case of 
classical {\it or} quantum descriptions, but, instead, classical 
{\it and} quantum descriptions. In other words, {\it classical and 
quantum formulations should coexist}.

The first part of the present article is devoted to a picture of classical 
and quantum physics fully consistent with the view that both formulations 
coexist. The second part of this article discusses the meaning of the 
variables used in phase space path integrals of different types, and 
shows how quantum mechanics impacts on the meaning of phase space 
variables that are used in classical mechanics. In the third part of 
this article, we show how the addition of a metric to the phase space of
classical mechanics can be considered the key concept in defining the 
basic ingredients in quantum mechanics, and in particular the impact that 
the addition of the metric to the classical phase space has on the phase 
space of quantum mechanics. 

Throughout this article we will rely on the concept of coherent states and 
their important application in building a bridge between the classical and 
the quantum world views. It is perhaps appropriate, therefore, that at this 
moment we present a mini review of the definition and some properties of 
coherent states before we put them to use in what follows.

\subsection{Coherent states: Definitions and properties}
Quantum mechanics deals with vectors, let us call them $|\psi\>$ following 
the notation of Dirac, in a Hilbert space $\frak H$, $|\psi\>\in{\frak H}$, 
with an inner product denoted by $\<\phi|\psi\>$ taken to be linear in the 
right hand vector and antilinear in the left hand vector. Linear operators 
act on vectors and yield new vectors, and in the quantum mechanics of a 
single canonical system, two basic operators are $P$ and $Q$, an 
irreducible pair,  which satisfy the Heisenberg commutation relation, 
$[P,Q]=-i\hbar\one$, where $\one$ denotes the unit operator. These 
operators are to be taken as self adjoint, which means that they can be 
used to generate unitary groups of transformations acting in the Hilbert 
space. In particular, we introduce the family of unitary operators given by
   \bn  U[p,q]\equiv e^{-iqP/\hbar}\,e^{ipQ/\hbar}\;,  \en
defined for all $(p,q)\in{\mathbb R}^2$, and which, we state, have the 
following multiplication rule
   \bn  U[p,q]\,U[p',q']=e^{i\s(pq'-qp')/2\hbar}\,U[p+p',q+q'] \;. \en

Next, we choose a distinguished vector $|\eta\>$, called the {\it fiducial 
vector}, 
which is normalized such that $\||\s\eta\>\s\|\equiv\sqrt{\<\eta|\eta\>}=1$, 
and consider the set of vectors each of which is defined as
  \bn  |p,q\>\equiv U[p,q]\,|\eta\>  \en
for all $(p,q)\in{\mathbb R}^2$. As defined, every vector in this set has 
a unit norm, $\|\s|p,q\>\s\|=1$. The set of states defined in this manner 
constitutes {\it a set of coherent states}; see, e.g., \cite{klabs}. 

Any set of coherent states as defined above satisfies two fundamentally 
important properties. The first property is that {\it the coherent state 
vectors are continuously parameterized}. This means that if the parameters 
$(p',q')\ra(p,q)$ in the sense of convergence in ${\mathbb R}^2$, i.e., 
$|p'-p|+|q'-q|\ra0$, then it follows that the vectors $|p',q'\>\ra|p,q\>$ 
in the strong sense, that is $\|\s|p',q'\>-|p,q\>\s\|\ra0$. The second 
property is that the vectors not only span the full Hilbert space, but 
that {\it they admit a resolution of unity as an integral over one 
dimensional projection operators} given by
  \bn  \one=\int|p,q\>\<p,q|\,dp\, dq/(2\pi\hbar)\;. \en
Here the integral runs over all points $(p,q)\in{\mathbb R}^2$, and the 
integral converges in the weak sense (as well as in the strong sense).

As minimum requirements in the choice of $|\eta\>$ it is generally 
convenient to require that
  \bn  \<\eta|P|\eta\>=0\;,\hskip1cm \<\eta|Q|\eta\>=0\;,  \label{h5}\en
which has the virtue that
  \bn  \<p,q|P|p,q\>=p\;,\hskip1cm \<p,q|Q|p,q\>=q\;, \en
leading to the physical interpretation of the variables $p$ and $q$ as 
{\it expectation values} in the coherent states. It is furthermore common 
to choose $|\eta\>=|0\>$, the ground state of an harmonic oscillator, 
typically satisfying the relation $(Q+iP)\s|0\>=0$. 

\section{Equations of Motion, and Boundary Data}
Classical mechanics is described by time dependent phase space paths, $q(t)$ 
and $p(t)$, where $q$ denotes position and $p$ denotes momentum. The 
equations that govern the time dependence may be determined as the 
extremal equations---also known as the Euler-Lagrange equations---that 
arise from stationary variation of the action functional
  \bn I=\tint[p\s{\dot q}-H(p,q)]\,dt  \en
holding $q$ fixed at both the initial and final times, say, $t=0$ and 
$t=T$, respectively. In this expression, $H(p,q)$ denotes the all-important 
Hamiltonian function, which, for convenience, we have assumed to be time 
independent. Variation of the action leads to
  \bn \delta I=\tint[\s({\dot q}-\d H/\d p)\s\delta p-({\dot p}+
\d H/\d q)\s\delta q\s]\,dt  \en
since the surface term, $p\s\delta q|$, vanishes. Setting $\delta I=0$ 
for general variations $\delta p(t)$ and $\delta q(t)$, we arrive at 
Hamilton's equations of motion
 \bn {\dot q}=\frac{\d H}{\d p}\;,\hskip1cm{\dot p}=-\frac{\d H}{\d q}\;. \en

How these equations are derived and whether or not they have solutions are 
{\it two fundamentally different issues}. In other words, holding 
$q$ fixed at two different times, as was used in this derivation, does 
{\it not} imply there is necessarily one and only one solution to these 
equations. Sometimes there may be a unique solution, but there also may 
be {\it no} solution, or even {\it many} solutions. As examples, consider 
${\dot q}= p$ and ${\dot p}=-q$, $q(0)=0$ and $q(2\pi)=1$ 
({\it no} solution) or $q(2\pi)=0$ ({\it many} solutions). In higher 
dimensions, there may be no solution for a path that needs to pass through 
a wall, etc. 

Accepting the difference between the derivation and the solution of the 
equations of motion allows us to consider more general situations. 
Suppose the action functional had the form
  \bn I'=\tint[\half(p{\dot q}-q{\dot p})-H(p,q)]\,dt  \en
which differs by a total derivative from the previous expression, 
specifically
  \bn  I'=I -\half\tint [d(pq)/dt]\,dt \;.  \en
We now propose to derive the same basic equations of motion from $I'$ that 
we obtained from $I$. To do so requires that we hold both $q$ 
{\it and} $p$ fixed at the beginning and end of the time interval in 
order that, after integration by parts, we again find that
  \bn \delta I'=\tint[\s({\dot q}-\d H/\d p)\s\delta p-({\dot p}+
\d H/\d q)\s\delta q\s]\,dt\;.  \en
Insisting that $\delta I'=0$ leads again to the equations of motion
  \bn {\dot q}=\frac{\d H}{\d p}\;,\hskip1cm{\dot p}=-\frac{\d H}{\d q}\;. \en
These equations are the same as before, but now we are asked to seek 
solutions based on the boundary conditions $p(0),q(0)=p',q'$ and 
$p(T),q(T)=p'',q''$. Clearly, in the general case the proposed solution 
is over specified and does not exist. However, in the present case---and 
unlike the previous case---we know how to limit the specified data so as 
to ensure a {\it unique} solution. One rule to obtain a unique solution 
would be to choose only the initial data $p(0),q(0)$, while another 
choice would be to choose only the final data $p(T),q(T)$.

\section{On the Meaning of the Variables $p$ and $q$}
In the phase space path integral formulation of quantum mechanics, 
Feynman \cite{fey} states that we may compute the propagator according 
to the formula
 \bn \<q''|\s e^{-i\H T/\hbar}\s|q'\>\equiv {\cal M}\int 
e^{(i/\hbar)\tint[p\s{\dot q}-H(p,q)]\,dt}\;\D p\, \D q\;.  \label{e12}\en
Here one is instructed to integrate over all functions $p(t)$ and 
$q(t)$, $0<t<T$, subject only to the conditions that $q(T)=q''$ and 
$q(0)=q'$. 
For present purposes it is not too important just how the integral on 
the right side is to be defined; rather we are more interested in the 
simpler question of what is the {\it meaning} of $q$. The meaning of the 
variable $q$ can be determined from the eigenvalue equation 
$Q|q''\>=q''|q''\>$, which asserts that $q$ has the meaning of the 
sharp value of position associated with an eigenvalue. We can determine 
the meaning of the variable $p$ by using a related phase space path 
integral given by
  \bn \<p''|\s e^{-i\H T/\hbar}\s|p'\>\equiv{\cal M}\int 
e^{(i/\hbar)\tint[-q\s{\dot p}-H(p,q)]\,dt}\;\D p\,\D q\;.  \label{e13}\en
In this case the integrals run over all functions $q(t)$ and $p(t)$, 
$0<t<T$, subject only to the conditions that $p(T)=p''$ and $p(0)=p'$. 
Consequently, the meaning of $p$ is that of a sharp eigenvalue associated 
with the equation $P|p''\>=p''|p''\>$.

It may seem reasonable that both $p$ and $q$ have the meaning of sharp 
values until one realizes that this situation refers to 
{\it quantum mechanics} and not {\it classical mechanics} since 
$\hbar\not=0$. In fact, this interpretation asserts that we can specify 
both $p$ {\it and} $q$ {\it simultaneously} for {\it all} intermediate 
$t$ values, while the uncertainty relation asserts that this is impossible 
for {\it any} $t$ value. 

It is interesting to observe that there is another interpretation of 
such phase space path integrals \cite{kla60,klabs}, namely, that
  \bn \<p'',q''|\s e^{-i\H T/\hbar}\s|p',q'\>\equiv {\cal M}\int 
e^{(i/\hbar)\tint[p\s{\dot q}-H(p,q)]\,dt}\;\D p\,\D q\;.   \label{e14}\en
Here, unlike the previous cases, one integrates over all $p(t)$ and all 
$q(t)$,
$0<t<T$, subject to the conditions $p(T),q(T)=p'',q''$ and $p(0),q(0)=p',q'$.
The initial and final eigenstates in this case are {\it not} sharp 
eigenstates but rather are the {\it coherent states} $|p,q\>$ discussed 
above for a general $|\eta\>$ which satisfies (\ref{h5}). For coherent 
states, the meaning of $p$ and $q$ is {\it not} that of sharp eigenvalues 
but rather is that of {\it mean} values, as already noted previously, 
namely,
  \bn  \<p,q|P|p,q\>=p\;,\hskip1cm \<p,q|Q|p,q\>=q\;. \en
As mean values, it is perfectly acceptable to specify values of $p(t)$ 
and $q(t)$ simultaneously for all $t$, $0<t<T$, and thus there is absolutely 
no contradiction with the uncertainty relation.

One should wonder how it is that the same formal path integral has two 
different evaluations; cf.~Eqs.~(\ref{e12}) and (\ref{e14}). The key word 
in the previous sentence is ``formal'', which implies that the so-called 
path integrals are in fact undefined as they stand and these formal 
expressions need to be {\it defined}. That the results given in 
(\ref{e12}) and (\ref{e14}) are different, means that the individual 
formal expressions have received their definition by different rules. 
These matters are well spelled out elsewhere \cite{kla13} and need not be 
repeated here.

The conclusion of the discussion in this section is that the only 
reasonable interpretation of the variables $p$ and $q$ is as mean values 
rather than truly sharp values since we live in a world where 
$\hbar\not=0$. This interpretation has important implications and is 
not changed by any attempt to define (\ref{e12}) or (\ref{e13}) by 
lattice limits as are customarily used to give some level of proper 
definition to such expressions \cite{swe}. 

\section{Shadow Metric}
Classical mechanics takes place on a phase space manifold which is 
equipped with a {\it symplectic form} $\omega$ leading to a 
{\it symplectic manifold}. The symplectic form takes as its argument 
two vectors and returns a number. For example, 
  \bn \omega(dp,dq)=\omega_{ab}\s dp^a\s dq^b\;,  \en
where $\omega_{ba}=-\omega_{ab}$ and $\det[\omega_{ab}]\not=0$. A 
symplectic manifold has what is called a 
{\it symplectic geometry} induced by the symplectic form. Such a 
geometry is rather
loose, like that of a rubber sheet which may be stretched by different 
amounts in different directions and still retain its ``geometry''. This 
kind of geometry is quite distinct from the more familiar Riemannian 
geometry determined by a Riemannian metric, and it is basic that a 
symplectic manifold appropriate for classical mechanics should not be 
assumed to be a metric space endowed with a Riemannian metric.

Our point of view is that one nevertheless {\it needs} a metric in order 
to give physics to the usual mathematical expressions that appear in 
classical mechanics \cite{kla77}. For example, one can make canonical 
coordinate transformations such that the mathematical expression for the 
Hamiltonian assumes the simple form $p$ for essentially any system. How 
is one to read out of the universal expression $H=p$ that the given 
expression actually refers to, say, an oscillator, or perhaps an 
anharmonic oscillator, etc? Clearly, one needs more information to make 
that choice correctly. And make no mistake, one definitely needs to make 
that choice because in quantum mechanics one solves for the spectrum of an 
oscillator or an anharmonic oscillator, etc., and no ambiguity in that 
situation can be admitted. To quantize a given classical theory one must 
not only know its mathematical formulation in terms of the phase space 
variables, but, at the same time, one must know to what physical system 
it refers. Unfortunately, this kind of information is simply {\it not} 
part of a traditional classical theory.

We must then seriously consider augmenting traditional classical mechanics 
with another structure the purpose of which is to keep track of the physics 
of the various mathematical expressions that enter into the theory. The 
structure we propose is that of a Riemannian metric, generally having the form
  \bn  d\sigma^2=A(p,q)\s dp^2+2B(p,q)\s dp\s dq+C(p,q)\s dq^2\;,  \en
with $A>0$, $C>0$, and $AC>B^2$. 
For the sake of illustration, let us choose the simplest example given by
  \bn  d\sigma^2=dp^2+dq^2 \;,  \en
which by its very form describes a two dimensional flat space expressed in
Cartesian coordinates. This metric does not enter into the analysis of 
classical mechanics; specifically, it does not enter at all into the 
classical equations of motion. Its sole purpose is ``to stand on the 
sidelines'' and to determine the physics of any given expression. How do 
we intend to do this? Let us adopt the expression $\half(p^2+q^2)+
\lambda q^4$ to correspond to the physics of an anharmonic oscillator 
provided that in the same coordinates the metric reads 
$d\sigma^2=dp^2+dq^2$. As the coordinates change, the expression for the 
Hamiltonian of this system changes form, but so does the expression for 
the metric! To understand the physical meaning of the new expression for 
the Hamiltonian, one need only find the transformation that restores the 
metric to Cartesian form and---presto!---the Hamiltonian in these 
coordinates assumes its usual physical form. In short, {\it the metric 
encodes the physical meaning of the mathematical expression that 
represents the Hamiltonian}. For the classical theory, this metric is 
separate from the theory itself, and for that reason it has been called 
a ``shadow metric'', one that is not part of the necessary mathematical 
formalism. However, as we have just argued, it is needed so as to complete 
the whole theory by ensuring that the physical meaning of the given 
expressions are not lost. In point of fact---and perhaps without even 
being aware of it---one intuitively introduces a ``shadow metric'', or 
its equivalent, so one can maintain a proper physical interpretation of 
the expressions that appear; otherwise the mathematics would be 
disconnected from the physics, a situation that would be untenable. 

\section{Continuous Time Regularization}
As remarked earlier, path integrals in general, and phase space path 
integrals in particular, are formal expressions that need some form of 
regularization in order for them to become defined. The most common 
regularization is a so-called lattice regularization which replaces the 
action integral by a Riemann sum and integrates over the path values at 
the discrete times steps that are thereby introduced. The result of 
interest arises in the limit that the lattice spacing goes to zero, a 
limit which is taken as the final step in the calculation. Of course, 
there are many different possible choices that one can make for the discrete 
form of the classical action all of which yield the usual classical action 
for continuous and differentiable paths in the limit that the lattice 
spacing vanishes. However, and this is the important point, there is no 
guarantee that the limit of the integral exists as the lattice spacing 
vanishes, and even when it exists, there is no guarantee that the result 
is physically acceptable for any of the various different choices of 
lattice action. This is an issue of great importance that needs to be 
analyzed whenever one is in any doubt about such problems.

On the other hand, a lattice form of regularization is not the only choice 
that can be used, and in this section we wish to discuss a form of 
regularization that maintains the classical action as an integral over 
time, and which is called a {\it continuous time regularization} to 
distinguish it from the lattice form previously discussed. The key idea 
here is the introduction of an additional factor that serves as a 
regularization device. This regularization factor---and its removal---takes 
the form \cite{dau}
  \bn \lim_{\nu\ra\infty}\s{\cal M}\int e^{(i/\hbar)
\tint[p\s{\dot q}-H(p,q)]\,dt}\;
e^{-(1/2\nu)\tint[{\dot p}^2+{\dot q}^2]\,dt}\;\D p\,\D q\;.  \label{g21}\en
Observe what has been done: We have inserted a real damping factor in the 
integrand that formally tends to unity as $\nu\ra\infty$. But, that limit 
is reserved until the integral over paths has actually been performed. It 
is important to observe that the regularized expression (\ref{g21}) can 
actually be given an unambiguous mathematical version as
  \bn \lim_{\nu\ra\infty}2\pi\s e^{\nu T/2\hbar}\int 
e^{(i/\hbar)\tint[p\s dq-H(p,q)\s dt]}\;d\mu^\nu_W(p,q)\;, \label{f25}\en
where $\mu^\nu_W$ denotes a Wiener measure on a flat two-dimensional 
phase space expressed in Cartesian coordinates and for which the parameter 
$\nu$ denotes the diffusion constant. Note well that this expression 
has {\it no} formal prefactor, and with $\tint p\s dq$ interpreted as a 
(Stratonovich) stochastic integral \cite{jjj}, (\ref{f25}) defines, for 
each $\nu<\infty$, a completely unambiguous path integral over continuous 
phase space trajectories. Moreover, it can be shown \cite{dau} for a wide 
class of Hamiltonians that the limit exists and that the limit actually 
provides a solution to Schr\"odinger's equation.  

\section{Existence of a Flat Space Metric in Any Canonical Quantum Theory}
There are several standard rules of quantization---such as those of 
Heisenberg and Schr\"odinger---that do not explicitly use a metric on a 
flat phase space in their construction. On the other hand, the fact that 
such rules do not use such a flat space metric does not mean that there 
is no such metric. In fact, perhaps surprisingly, there is {\it always} 
such a metric implicitly present, as we now proceed to demonstrate. Everyone 
would agree that any canonical quantization procedure leads to vectors in a 
Hilbert space and to canonical self adjoint operators $Q$ and $P$ that 
satisfy the Heisenberg commutation relation, $[Q,P]=i\hbar\one$. With the 
aid of these basic elements, we can always construct vectors of the form
  \bn  |p,q;\psi\>\equiv e^{-iqP/\hbar}\,e^{ipQ/\hbar}\,|\psi\>\;,  \en
for a general vector $|\psi\>$, in whatever representation is involved. In 
terms of $d\s|p,q;\psi\>\equiv|p+dp,q+dq;\psi\>-|p,q;\psi\>$, we next form 
the expression
  \bn  \|\s\hbar\s d|p,q;\psi\>\s\|^2-
|\<p,q;\psi|\s\hbar\s d\s |p,q;\psi\>|^2 \;, \label{h24}\en
which is evidently quadratic in the differentials $dp$ and $dq$. 
Granting minimal domain requirements, it readily follows that (\ref{h24}) 
becomes
 \bn  \<(\De Q)^2\>\, dp^2+\<(\De P\De Q+\De Q\De P)\>\, dp\s dq+
\<(\De P)^2\>\,dq^2\;,  \label{g17}\en
where in this expression, we have used the notation 
  $\<(\cdot)\>\equiv\<p,q;\psi|(\cdot)|p,q;\psi\>$, and 
$\De Q\equiv Q-\<Q\>$, etc. It is clear that the given expression 
generates a metric on phase space, and since the metric coefficients 
$\<(\De Q)^2\>$, etc.,  are constants, then the metric characterizes a 
{\it flat} two dimensional phase space. This feature is embedded into 
any version of canonical quantization one can imagine. A flat space 
metric may not be explicitly {\it used} in arriving at your favorite 
quantization, but it is nevertheless present in the multitude of 
consequences that follow from the quantization itself. 

Moreover, for many states of physical interest, e.g., a typical harmonic 
oscillator ground state, it follows that the coefficients appearing in 
(\ref{g17}) are all proportional to $\hbar$. Thus, as seems entirely 
natural, the phase space metric induced by any quantization procedure is 
fully of a quantum character itself.

\section{Metrical Quantization}
It is clear that one may sometimes choose a different set of axioms to 
derive a common body of knowledge, and that dictum certainly applies to 
the process of quantization, as exemplified by the approaches of 
Heisenberg, Schr\"odinger, and Feynman. We have observed that 
traditional canonical quantization procedures do not make use of a flat 
phase space metric, but that such a metric nevertheless arises in a 
natural way within any quantization scheme. Suppose we took that metric, 
normally a secondary feature, and promoted it to primary status, namely 
postulating its existence as one of our axioms of quantization, indeed, 
as our {\it primary} axiom of quantization. The existence of a flat phase 
space metric and its elevation to prominence has another advantage for, 
as we have observed above, a metric we have called the ``shadow metric'' 
is an all but essential ingredient in order to keep track of the physics 
of the mathematical expressions under consideration. 

Thus as our first axiom in the program of metrical quantization for 
canonical systems, we postulate the existence of a flat space metric 
$d\sigma(p,q)^2$ on our classical phase space which for the sake of the 
present discussion we take in the form
  \bn  d\sigma(p,q)^2=\hbar(dp^2+dq^2)\;,  \label{g18}\en
a relation that evidently characterizes a flat two dimensional phase 
space expressed in Cartesian coordinates.
Observe that this is a quantum object since it is proportional to $\hbar$. 
In the formal classical limit in which $\hbar\ra0$, it follows that the 
metric vanishes; this property ensures that the metric is strictly quantum 
in character and not part of the usual classical theory as conventionally 
interpreted. On the other hand, since $\hbar>0$ in the real world, 
we can still use (\ref{g18}) as our shadow metric to give physics to our 
expressions. 

Quantum mechanics {\it needs to know} what system is under discussion, it 
needs to know so that it can determine the energy spectrum, for example, 
of one specific system and not that of any other system. Thus it seems 
absolutely natural that we must combine the metric directly with the 
quantization formalism so that in fact the quantization will ``know'' 
what system is under consideration. 

In an earlier section we discussed a continuous time regularization as a 
mathematical device to provide a regularization to an otherwise ill 
defined formal phase space path integral. Now, as we look at that 
expression once again, we see that the regularization itself involved the 
very same flat phase space metric we have in mind. That is, the given 
regularization accomplished the further goal of adding to the formalism an 
appropriate metric as the ``keeper of the physics''. In short, {\it not 
only does the continuous time regularization of the phase space path 
integral presented in (\ref{g21}) or (\ref{f25}) solve the issue of giving 
a proper mathematical meaning to the path integral, it {\bf simultaneously} 
solves the problem of maintaining a proper physical interpretation 
throughout the quantization procedure.} 

With this discussion as background, we are in a position to formalize the 
two axioms of Metrical Quantization \cite{swe}. First we adopt the phase 
space metric (\ref{g18}). The second, and final step, is to say how we 
use this metric and that is to postulate that the propagator is defined by 
  \bn \<p'',q''|\s e^{-i\H T/\hbar}\s|p',q'\>\equiv
\lim_{\nu\ra\infty}2\pi\s e^{\nu T/2\hbar}\int 
e^{(i/\hbar)\tint[p\s dq-H(p,q)\s dt]}\;d\mu^\nu_W(p,q)\;, \label{h35}\en
where the Brownian motion paths are carried by the metric (\ref{g18}) in 
the sense of (\ref{g21}).
This is all that is needed since, according to the Gel'fand-Naimark-Segal 
reconstruction theorem \cite{emch}, everything quantum in character about 
(\ref{h35}) is a consequence of the functional form of the right hand side. 
In particular, and as the notation suggests, it follows that the right hand 
side generates the propagator in a canonical coherent state representation 
based on self adjoint Heisenberg operators $Q$ and $P$ that satisfy 
$[Q,P]=i\hbar\one$. Morevover, in the present case it follows that the 
fiducial vector necessarily satisfies the relation $(Q+iP)\s|0\>=0$. 
Additionally, it also follows that the Hamiltonian operator $\H$ is 
related to the classical Hamiltonian $H(p,q)$ by 
{\it antinormal ordering\/}; that is, the very act of introducing the 
regularization and making the formal phase space path integral well 
defined has removed all ambiguity including the usual factor ordering 
ambiguity. In short, the very form of the regularization has already 
implicitly {\it chosen} an ordering prescription for the quantization!

\section{Covariance under Canonical Coordinate \\Transformations}
The fact that (\ref{h35}) [or more loosely, (\ref{g21})] is well defined 
as an integral means that we are free to change the variables of 
integration as usual. Let us discuss some of the changes we might make 
and see how they effect the form of the integrand. 

{}From a classical point of view, recall that a classical canonical 
transformation associates the old and new canonical coordinates via a 
one form such as
  \bn  {\o p}\,d{\o q}=p\s dq+dF({\o q},q)\;,  \en
where ${\o p},{\o q}$ denote the new canonical coordinates while $p,q$ 
denote the old canonical coordinates. Furthermore, the Hamiltonian 
transforms as a scalar in classical mechanics, which means that
  \bn  {\o H}({\o p},{\o q})\equiv H(p({\o p},{\o q}),q({\o p},{\o q}))=
H(p,q)\;. \en
In classical mechanics, such expressions make sense because one is dealing 
with smooth functions of time that are both continuous and have continuous 
derivatives. 

However, in the well defined path integral of (\ref{h35}) we are dealing 
with Brownian motion paths $p(t),q(t)$ which are continuous paths but 
which are {\it nowhere} differentiable! This fact means that expressions 
like $\tint p\s dq$ cannot be defined as ordinary integrals, but rather 
they should be interpreted as {\it stochastic integrals} \cite{jjj}. For 
the case at hand, we may adopt a Riemann sum prescription based on the 
mid-point rule of definition, namely
  \bn \tint p\s dq\equiv \lim\Sigma\,\half\s(p_{l+1}+p_l)(q_{l+1}-q_l) 
\;,  \en
in the limit that the lattice spacing goes to zero. This is the so-called 
Stratonovich prescription, which has the virtue that it satisfies the 
rules of the ordinary calculus \cite{jjj} despite the fact that the 
paths $p(t)$ and $q(t)$ are nowhere differentiable!\footnote{The 
commonly used alternative rule, which is the so-called It\^o prescription, 
and given by $\lim\Sigma\, p_l(q_{l+1}-q_l)$, has other virtues but it 
generally does not satisfy the ordinary rules of calculus. Due to our 
initial choice of coordinates, we are free to choose either the 
Stratonovich or the It\^o rule with which to make our coordinate 
transformations since both rules lead to the same result in Cartesian 
coordinates for the integral in question.} 

This correspondence with the ordinary rules of calculus means that in 
the continuous time regularized path integral the classical action will 
transform just as it does in the classical theory. The Wiener measure 
will also undergo a coordinate transformation, but it will still 
describe Brownian motion on a {\it flat} two dimensional phase space, 
although in general it will now do so in curvilinear coordinates rather 
than the initial Cartesian coordinates. Therefore under a canonical 
coordinate transformation
(\ref{h35}) becomes \cite{dau,kla32}
  \bn \<{\o p}'',{\o q}''|\s e^{-i\H T/\hbar}\s|{\o p}',{\o q}'\>=
\lim_{\nu\ra\infty}{\cal M}\int e^{(i/\hbar)\tint[{\o p}\s d{\o q}+
d{\o G}({\o p},{\o q})-{\o H}({\o p},{\o q})\s dt]}\;
d{\o\mu}^\nu_W({\o p},{\o q})\;. \en
In this expression, the function ${\o G}({\o p},{\o q})$ [which arises 
from $F({\o q},q)$] appears as a total derivative and so amounts, in 
effect, to the addition of a phase factor to each of the coherent states. 

Here, at last, we can see the virtue of the present formulation very 
clearly. Although the coordinate form of the Hamiltonian may well have 
changed, and thus the physical meaning of the Hamiltonian cannot be read 
directly from its functional form in the non-Cartesian coordinates, the 
phase space path integral nevertheless still refers to the original 
physical system and what ensures this is the fact that the metric on 
flat space needed to carry the Brownian motion paths acts as a shadow 
metric maintaining control over the proper physical interpretation. 
Moreover, we can also see that the quantum Hamiltonian $\H$ has in no 
way changed so that indeed the original physical system is still under 
discussion. Observe also that the result is still expressed in terms of 
the original coherent states---apart from a trivial phase factor---the 
only difference being the way in which they are parametrized. In particular,
  \bn |{\o p},{\o q}\>\equiv e^{-i\s {\o G}({\o p},{\o q})/\hbar}\,
e^{-i\s q({\o p},{\o q})\s P/\hbar}\,e^{i\s p({\o p},{\o q})\s 
Q/\hbar}\,|0\>\;. \en 
These states still enjoy a resolution of unity, but which is now 
expressed in the form
  \bn  \one=\int |{\o p},{\o q}\>\<{\o p},{\o q}|\,d{\o p}\s 
d{\o q}/(2\pi\hbar)\;. \en

In summary, and in a manor of speaking, our main point is that quantum 
phase space may be said to arise from classical phase space simply by 
the introduction of a metric. Thus, viewed in the right way, perhaps 
classical and quantum mechanics are closer to each other than is 
commonly believed!

\end{document}